\documentclass[a4paper]{article}

\usepackage{INTERSPEECH2022}
\usepackage{multirow}
\usepackage{enumitem}

\title{Characterizing Therapist's Speaking Style in Relation to Empathy in Psychotherapy}
\name{Dehua Tao$^1$, Tan Lee$^1$, Harold Chui$^2$, Sarah Luk$^2$}
\address{
  $^1$ Department of Electronic Engineering \quad 
  $^2$ Department of Educational Psychology\\The Chinese University of Hong Kong}
\email{dhtao@link.cuhk.edu.hk, tanlee@ee.cuhk.edu.hk, \{haroldchui, sarah\_luk\}@cuhk.edu.hk}

\begin{document}

\maketitle
\begin{abstract}
In conversation-based psychotherapy, therapists use verbal techniques to help clients express thoughts and feelings, and change behavior. In particular, how well therapists convey empathy is an essential quality index of psychotherapy sessions and is associated with psychotherapy outcome. In this paper, we analyze the prosody of therapist speech and attempt to associate the therapist's speaking style with subjectively perceived empathy. An automatic speech and text processing system is developed to segment long recordings of psychotherapy sessions into pause-delimited utterances with text transcriptions. Data-driven clustering is applied to the utterances from different therapists in multiple sessions. For each cluster, a typological representation of utterance genre is derived based on quantized prosodic feature parameters. Prominent speaking styles of the therapist can be observed and interpreted from salient utterance genres that are correlated with empathy. Using the salient utterance genres, an accuracy of $71\%$ is achieved in classifying psychotherapy sessions into ``high" and ``low" empathy level. Analysis of results suggests that empathy level tends to be (1) low if therapists speak long utterances slowly or speak short utterances quickly; and (2) high if therapists talk to clients with a steady tone and volume.
\end{abstract}

\section{Introduction}
\label{sec:intro}

Psychotherapy is a kind of treatment or intervention process that aims to help people express emotion, change behavior and overcome difficulties in the desired way. It typically takes the form of spoken conversation between a therapist and a client. Psychotherapy is an asymmetric help-relationship focused on the client. The therapist would have a main and deterministic contribution to the treatment outcomes. Ackerman {\it et al.} reviewed therapist's characteristics and techniques that positively \cite{ackerman2003review} and negatively \cite{ackerman2001review} influenced psychotherapy outcomes. They found that therapist's personal attributes, such as being honest, respectful, and interested, and therapist's techniques, such as exploration and reflection, positively impacted the therapeutic alliance \cite{ackerman2003review}. In another study \cite{ackerman2001review}, therapist's personal attributes, such as being rigid, uncertain, and distant, and therapist's techniques, such as unyielding use of transference interpretation and inappropriate use of silence, were found to contribute negatively to the therapeutic alliance.

During a psychotherapy session, the therapist and the client conduct a specific type of communication through therapeutic conversation. Speech is the primary medium in the conversation for conveying messages between speakers. As a kind of speaker-specific characteristic, speaking style \cite{eskenazi1993trends, mokhtari2008speaking, grimaldi2009speech} of the therapist, i.e., ``how one says", is believed to be associated with the psychotherapy outcome. The speaking style is considered to be closely related to speech prosody, i.e., intonation, stress, and rhythm \cite{llisterri1992speaking, abe1997speaking, sonmez1998modeling, adami2003modeling}.

Therapist empathy is an essential quality index in psychotherapy. Empathy is described as ``the therapist's sensitive ability and willingness to understand the client's thoughts, feelings, and struggles from the client's point of view" \cite{rogers1995way}. High scores of therapist empathy are correlated with treatment retention as well as positive clinical outcomes \cite{elliott2011empathy, miller2009toward}. Therefore, empathy rating, namely the subjectively rated empathy level, is considered a useful index to guide the search of desired speaking style of the therapist, i.e., to determine how the therapist's speaking style is correlated, positively or negatively, with empathy.

In this work, therapist's speaking style is quantified based on prosodic cues in speech. Specifically, automatically computed parameters of duration, pitch and intensity are used as the bases for feature extraction. Previous studies revealed that prosody and empathy are closely related. The brain regions responsible for prosody production and perception are also utilized for aspects of empathy \cite{aziz2010common}. The study by Regenbogen {\it et al.} showed that behavioral empathy in human communication relied on consistent information from three channels, i.e., facial expression, prosody and speech content \cite{regenbogen2012differential}. Xiao {\it et al.} studied prosodic correlates of empathy by quantizing prosodic features and estimating the joint distribution of certain prosodic patterns. Their results suggested that high pitch and energy of therapist speech were negatively correlated with empathy \cite{xiao2014modeling}. In \cite{imel2014association}, it was shown that the correlation between therapist's and standardized patient's mean pitch values was higher in high-empathy sessions compared to low-empathy ones. Speech rate was investigated in \cite{xiao2015analyzing}, in which the mean absolute difference of turn-level speech rates between therapists and clients was found to correlate with therapist empathy.

The paper first presents an automatic speech and text processing system for segmenting long recordings of psychotherapy sessions into pause-delimited utterances that are aligned with text transcriptions. The speech utterances from different therapists in multiple sessions are analyzed by clustering. The notion of utterance genre is introduced as an interpretable representation of utterance-level prosodic characteristics in a cluster. The correlation between session-wise contribution ratio of utterance genres and empathy ratings is analyzed to identify utterance genres that have significant effect on empathy. The therapist's speaking style can be explained and summarized in terms of utterance genres. Based on the utterance genres, a psychotherapy session can be automatically classified as having "high" or "low' empathy level.


\section{Psychotherapy speech database}
\label{sec:database}

\subsection{Speech data}
\label{ssec:sphdata}


Speech data being used in this study is a part of recordings from counseling practicums for therapist trainees at the Chinese University of Hong Kong. Clients of counseling were adults who sought counseling assistance concerning career, relationship, emotion, and stress. The study was approved by the institutional review board, and informed consent was obtained from both the clients and therapists. The speech data consists of a total of 156 psychotherapy sessions from 39 different pairs of therapists and clients, i.e., each therapist-client pair has 4 sessions. All therapists and clients spoke Hong Kong Cantonese, with occasional English code-mixing. Each session is about 50 minutes long. The therapist and the client each wore a lavalier microphone clipped between the collar and chest during the session. The 156 sessions of recordings were manually transcribed. In each session, a therapist and a client took turn to speak. A speaker turn refers to the time period in which there is only one active speaker. Speech transcription was done on speaker turn basis, with the speaker identity (therapist or client) marked and the speech content transcribed into Chinese characters with punctuation marks.

Subjective rating was performed on each session using the Therapist Empathy Scale (TES) \cite{decker2014development}. The TES is a nine-item observer-rated measure of therapist empathy. Each item is rated on a seven-point scale from 1 $=$ \textit{not at all} to 7 $=$ \textit{extremely}. The total score (range from 9 to 63) is used in this study, with a larger value indicating higher therapist empathy.


\subsection{Turn-level speech-text alignment}
\label{ssec:aligner}

The speech data of each speaker in a session needs to be extracted for analysis. However, there is no time stamp information of therapist and client speech in these recordings. With speaker-turn-based transcription available, it is feasible to find the beginning and ending time of all speaker turns over a 50-minute long recording.  

\begin{figure*}[htb]
\centering
\centerline{\includegraphics[width=16.5cm]{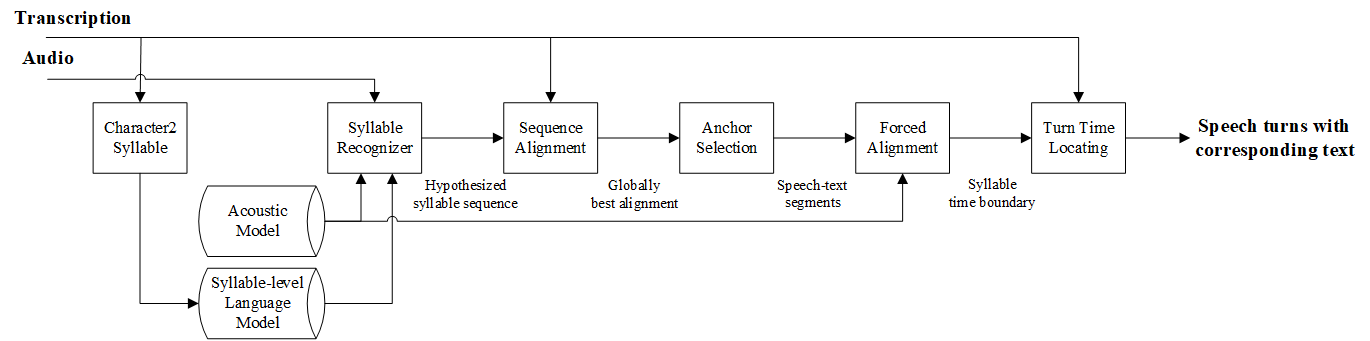}}
\vspace{-1em}
\caption{Flow diagram of the turn-level speech-text alignment system.}
\label{fig:aligner}
\vspace{-1.0em}
\end{figure*}

Inspired by \cite{moreno1998recursive}, an automatic system for turn-level speech-text alignment is developed. Figure \ref{fig:aligner} shows the block diagram of the alignment process. Given a long recording of psychotherapy session and its text transcription, automatic turn-level speech-text alignment is performed in the following steps.

\vspace{0.5em}
\noindent
\textbf{Syllable recognition with session-specific language model:} The \textit{Character2Syllable} module converts Chinese characters in the transcription into Cantonese syllables in the form of Jyutping symbols \cite{lee2002spoken}. Code-mixing English words are also converted into Jyutping symbols that represent their pronunciations as close as possible \cite{chan2009automatic}. The Cantonese syllable recognizer is used to produce a hypothesized sequence of syllables (in Jyutping symbols) for all speech utterances in the session. The syllable recognition is performed with a tri-gram language model trained with the syllable transcription of this session. Using such session-specific language model helps pruning irrelevant hypotheses in syllable recognition. Initially an acoustic model trained with other Cantonese databases is used to decode speech of all sessions. With the hypothesized syllable sequences, an in-domain acoustic model is trained as described below.

\vspace{0.5em}
\noindent
\textbf{Syllable alignment and anchor selection:} The hypothesized syllable sequence is aligned with the manual transcription. The globally best alignment is determined by applying dynamic programming. Aligned sub-sequences of syllables are selected as anchors. Each anchor is a long sequence of consecutive syllables, and believed to be correctly aligned. The anchors are used to partition the transcription and speech signal in the long session into short segments. Then an acoustic model is trained with the corresponding speech and text segments from all sessions. The new acoustic model would replace the existing one from which the hypothesized syllable sequence was obtained. The above steps are repeated and speech-text segments are updated. 

\vspace{0.5em}
\noindent
\textbf{Forced alignment and turn time locating:} Forced alignment on each speech-text segment is performed to obtain syllable-level time stamps, i.e., beginning and ending time of each syllable. The text content in the segments are then related to the manual transcription to locate the first and last syllables of each speaker turn.

\vspace{0.5em}
In this work, 118 counseling sessions involving 39 different therapists are used, with empathy ratings on the two extremes. The 118 sessions are divided into high-empathy and low-empathy sessions. Ratings of 61 high-empathy sessions range from 42 to 56.5 with the mean 46.34 \textpm 3.58, while 57 low-empathy sessions range from 18 to 36 with the mean 30.40 \textpm 4.79.




Forced alignment at syllable level is performed on each therapist turn. A long speech turn of therapist is divided into sub-turn utterances that are separated by major pauses of 0.5 second or longer. Utterances shorter than 0.5 seconds are not used in the analysis. Table \ref{tab:sphdata} gives a summary of the speech data used in the study.

\begin{table*}[htb]
\centering
\caption{Summary of the therapist speech data. There are totally 118 psychotherapy sessions involving 39 different therapists.} 
\label{tab:sphdata}
\vspace{-0.5em}
\resizebox{1.0\textwidth}{!}{
\begin{tabular}{|c|c|c|c|c|c|}
\hline
\begin{tabular}[c]{@{}c@{}}Average speech time \\ per session (min)\end{tabular} & \begin{tabular}[c]{@{}c@{}}Average no. of characters \\ per session\end{tabular} & \begin{tabular}[c]{@{}c@{}}Average no. of utterances \\ per session\end{tabular} & \begin{tabular}[c]{@{}c@{}}Average duration \\ of utterances (sec)\end{tabular} & \begin{tabular}[c]{@{}c@{}}Average no. of characters \\ per utterance\end{tabular} & \begin{tabular}[c]{@{}c@{}}Average empathy rating \\ of sessions\end{tabular} \\ \hline
12.74                                                                            & 3637                                                                             & 211                                                                              & 3.62                                                                            & 17                                                                                 & 38.64                                                                         \\ \hline
\end{tabular}
}
\vspace{-1.0em}
\end{table*}

\section{From utterance genre to speaking style}
\label{sec:uttge2spk}

\subsection{Feature extraction}
\label{ssec:featext}

Prosodic features are extracted from each sub-turn utterance. The features cover the aspects of time duration, pitch, and intensity of speech. Let $d$ denote the time duration of the utterance. In relation to the time proportion of speech and non-speech segments, speech rate of the utterance, denoted as $sr$, is defined as the number of Chinese characters (syllables) spoken in one second. Frame-level pitch and intensity values are computed using the script prosodyAcf.conf, one of the standard scripts in the openSMILE toolkit \cite{eyben2010opensmile}. The mean of frame-level pitch values over the utterance, is used to represent the utterance's overall pitch level. The standard deviation and the interquartile range are calculated to measure the degree of pitch variation over the utterance. These pitch-based feature parameters are denoted as $p_{\mu}$, $p_{std}$ and $p_{iqr}$ respectively. The same statistical parameters are computed for intensity and denoted as $i_{\mu}$, $i_{std}$, and $i_{iqr}$ respectively.

For each of the above utterance-level feature parameters, speaker normalization is applied to reduce inter-speaker variation. For example, let $p_{std}^T$ be the standard deviation of pitch computed from all utterances of the therapist in a session, and $p_{std}^u$ be the standard deviation of pitch of a specific utterance in this session. The normalized standard deviation of pitch of the utterance is obtained as $p_{std} = p_{std}^u / p_{std}^T$. The other utterance-level feature parameters are normalized in a similar way. As a result, a set of 8 normalized feature parameters, denoted as $(d, sr, p_{\mu}, p_{std}, p_{iqr}, i_{\mu}, i_{std}, i_{iqr})$, are used to represent one utterance.

\subsection{Utterance genre of cluster}
\label{ssec:uttge}

Clustering is commonly used to interpret unknown data and identify inherent clusters in the given feature space. In this work, clustering is applied to all sub-turn utterances of therapist speech in the 118 psychotherapy sessions. Different combinations of the 8 feature parameters are attempted and evaluated by experiments. For each combination, utterances from all sessions are split into a prescribed number of clusters. Each cluster contains utterances with similar feature values. In a sense utterances being put into the same cluster represent a specific utterance genre. To facilitate more intuitive interpretation of clustering results, we propose to use the feature pattern as suggested in \cite{xiao2014modeling} to serve as a typological representation of utterance genre. This is explained below using an example of two feature parameters $(d, sr)$.

\noindent
\textbf{Feature pattern:} Similar to the approach in \cite{xiao2014modeling}, each feature parameter is quantized into $Q$ equally populated intervals. The boundaries of the quantization intervals are determined based on all utterances of therapist speech in the 118 sessions. For example, with $Q = 3$, the feature parameter is quantized into discrete values according to the 33 and 67 percentiles. Verbally the quantized value can be interpreted as low ($L$), medium ($M$), and high ($H$) value. A \textbf{\emph{feature pattern}} for $(d, sr)$ is defined as a combination of two quantized values, e.g., $(L, L)$, or $(H, M)$. There are a total of $3^2=9$ different feature patterns for the combination $(d, sr)$.

\noindent
\textbf{Representation of utterance genre:} As mentioned earlier, utterances in the same cluster tend to have similar feature values. The majority of utterances in a cluster are expected to be quantized into one of feature patterns. Given a specific cluster, we count the number of utterances that are quantized into each of the feature patterns. The feature pattern with the largest number of utterances is used to represent the utterance genre of this cluster. In the case of two feature parameters $(d, sr)$ with $Q = 3$, an utterance genre represented by the feature pattern $(H, L)$ is described as utterances with long duration and low speech rate. The value range of the feature pattern is denoted as $([d_{min}^{H}, d_{max}^{H}], [sr_{min}^{L}, sr_{max}^{L}])$. The range of duration $H$ is obtained from the values of parameter $d$ of utterances quantized into the pattern $(H, L)$ in the cluster. The range of speech rate $L$ is obtained in the same way.

A specific utterance genre (cluster) may be found in many different sessions. However, its contribution varies from one session to another. Let $r_c^g$ denote the contribution ratio of the utterance genre $g$ of the cluster $c$ in a session. $r_c^g$ is obtained by counting the number of utterances carrying the genre $g$ in the cluster $c$ and dividing by the total number of utterances in the session. The Pearson's correlation $\rho$ between session-wise contribution ratio of the utterance genre $g$ and empathy ratings is computed. Through the analysis of correlation, utterance genres contributing positively or negatively to empathy can be identified. These utterance genres can be used to describe the speaking style of therapists in relation to empathy.

\section{Experiments and Results}
\label{sec:expres}

\subsection{Experimental setup}
\label{ssec:expdesign}

The $K$-means clustering algorithm is applied in our experiments. Multiple trials of clustering are attempted with different cluster number, i.e., $K = 2, 3 ..., N$. For the utterance genre $g$ of the cluster $c$ ($c = 1, 2, ..., K$), let $\bar{r}_c^g$ be its average contribution ratio over all sessions, and $pv_c^g$ denote the p-value of the correlation $\rho$ between its session-wise contribution ratio and empathy ratings. In a specific trial of clustering, the utterance genre $g$ is retained if it satisfies the following conditions: 

\begin{enumerate}[label=(\roman*)]
    \item In the cluster $c$, the occurrence ratio of the feature pattern representing the utterance genre $g$ is over a threshold $r^f_{min}$. The occurrence ratio is computed by counting the number of utterances in the feature pattern and dividing by the total number of utterances in the cluster.
    \item $\bar{r}_c^g$ is higher than a pre-set threshold $r^g_{min}$, meaning that an utterance genre is retained only if it is carried by a sufficient number of utterances.
    \item $pv_c^g$ is below a pre-determined threshold $pv_{max}$.
\end{enumerate}

The speaking style of therapists can be summarized from the retained, i.e., salient utterance genres. The values of $N$, $r^f_{min}$, $r^g_{min}$, and $pv_{max}$ are determined empirically to be 20, 0.5, 0.05, and 0.05 respectively in our experiments.

The 8 feature parameters described in Section \ref{ssec:featext} are divided into two groups: (a) $\{d, sr, p_{\mu}, i_{\mu}\}$ represent the mean values; (b) $\{(p_{std}, p_{iqr}), (i_{std}, i_{iqr})\}$ represent variation of feature values in the utterance. For each group, utterance clustering is performed on different combinations of feature parameters. The number of combinations for group (a) is 15 ($C_4^1 + C_4^2 + C_4^3 + C_4^4$), and for group (b) is 3 ($C_2^1 + C_2^2$). Thus there are a total of 18 combinations of feature parameters attempted in our experiments.

\subsection{Salient utterance genres}
\label{ssec:saluttge}

Table \ref{tab:patncorr} shows a list of salient utterance genres and the respective feature patterns in the case of $Q = 3$. Other choices of $Q$ will be discussed later in Section \ref{sec:disc}. The utterance genres in the table are either positively or negatively correlated with subjective ratings of therapist empathy. For example, Low intensity is positively correlated with therapist empathy, while short duration and low speech rate are negatively correlated. When therapists speak short utterances, they are not suggested to speak quickly. The analysis also shows that therapists express empathy better by speaking with a steady tone and volume. 

The relation between session-wise occurrence count of feature patterns and therapist empathy is examined in a similar way to that suggested in \cite{xiao2014modeling}. For each feature pattern from the 18 combinations of feature parameters, its session-wise occurrence ratio is computed by dividing the occurrence count by the total number of utterances in the session. Then the Pearson’s correlation between session-wise occurrence ratio and empathy ratings is computed. The feature patterns correlated with therapist empathy are found. Compared to the findings with the method of utterance genre, a similar trend is observed with the method of feature pattern distribution.


\begin{table}[htb]
\centering
\caption{Feature patterns of salient utterance genres for $Q = 3$: \\L - Low, M  - Medium, H - High}
\label{tab:patncorr}
\vspace{-0.5em}
\resizebox{1.0\linewidth}{!}{
\begin{tabular}{|c|c|c|c|}
\hline
\multicolumn{1}{|c|}{Feature pattern} & $\rho$ & p-value & Value range \\ \hline
$d$ = L \& $sr$ = H                   & -0.33  & 0.0003  & [0.155, 0.615] \& [1.208, 1.801] \\ \hline
$d$ = L                               & -0.28  & 0.002   & [0.364, 0.547] \\ \hline
$d$ = H \& $sr$ = L                   & -0.27  & 0.003   & [1.337, 2.096] \& [0.578, 0.883] \\ \hline
$sr$ = L                              & -0.21  & 0.03    & [0.558, 0.9] \\ \hline
$i_u$ = L                             & 0.20   & 0.03    & [0.853, 0.921] \\ \hline
$i_{std}$ = L \& $i_{iqr}$ = L        & 0.20   & 0.03    & [0.79, 0.87] \& [0.721, 0.895] \\ \hline
$p_{std}$ = L \& $p_{iqr}$ = L        & 0.19   & 0.04    & [0.64, 0.882] \& [0.041, 0.711] \\ \hline
\end{tabular}
}
\vspace{-1.5em}
\end{table}

\subsection{Binary classification of ``high" versus ``low" empathy}
\label{ssec:binaryclass}

After $N$ trials of clustering, it is found that utterance genres of several clusters are represented by the same feature pattern. The value range of the feature pattern representing such genres is computed. Assume there are $T$ clusters of which utterance genres are represented by the feature pattern $(H, L)$ in the case of two feature parameters $(d, sr)$ with $Q = 3$, the value range of the feature pattern, denoted as $([d_{\hat{min}}^{H}, d_{\hat{max}}^{H}], [sr_{\hat{min}}^{L}, sr_{\hat{max}}^{L}])$, is obtained in the following way: (1) the range in the cluster $t$ is computed by the way described in Section \ref{ssec:uttge}, denoted as $([d_{min}^{H}, d_{max}^{H}], [sr_{min}^{L}, sr_{max}^{L}])_t$, where $t = 1, 2, ..., T$; (2) $d_{\hat{min}}^{H}$ in the value range takes the median of $\{(d_{min}^{H})_1, (d_{min}^{H})_2, ..., (d_{min}^{H})_T\}$. $d_{\hat{max}}^{H}$, $sr_{\hat{min}}^{L}$ and $sr_{\hat{max}}^{L}$ are obtained in the same way. The value ranges of feature patterns of salient utterance genres for $Q = 3$ are given as in Table \ref{tab:patncorr}.

Now the contribution ratio of a salient utterance genre in each session can be obtained through the value range. The ratio is computed by counting the number of utterances with the feature values within the range and dividing by the total number of utterances in the session. Such contribution ratios of all salient utterance genres are used as the session-level features to predict ``high" or ``low" empathy level of a session. The dimension of the session-level features is equal to the number of salient utterance genres. 5-fold cross-validation is carried out. Linear Support Vector Machine (SVM) is used as the classifier. For comparison, the occurrence ratio of prominent feature patterns is used as the baseline method for the classification. Table \ref{tab:classaccu} shows the results of classification with different quantization intervals $Q$. The method of utterance genre performs better and achieves an accuracy of $71\%$ for $Q = 3$.

\begin{table}[htb]
\centering
\caption{Classification accuracy on sessions with high versus low level of therapist empathy}
\label{tab:classaccu}
\vspace{-0.5em}
\begin{tabular}{|c|c|c|c|}
\hline
Method & Q & \begin{tabular}[c]{@{}c@{}}\ Dimension of \\ session-level feature \end{tabular} & Accuracy \\ \hline
    \multirow{4}{*}{\begin{tabular}[c]{@{}c@{}}\ Contribution ratio of \\ utterance genres \end{tabular}}   & 2 & 29 & 0.66 \\ \cline{2-4} 
  & 3 & 25 & \textbf{0.71} \\ \cline{2-4} 
  & 4 & 16 & 0.67 \\ \cline{2-4} 
  & 5 & 13 & 0.67 \\ \hline
\multirow{4}{*}{\begin{tabular}[c]{@{}c@{}}\ Occurrence ratio of \\ feature patterns \end{tabular}}   & 2 & 17 & 0.60 \\ \cline{2-4} 
  & 3 & 21 & 0.62 \\ \cline{2-4} 
  & 4 & 19 & 0.69 \\ \cline{2-4} 
  & 5 & 12 & 0.66 \\ \hline
\end{tabular}
\vspace{-1.5em}
\end{table}

\section{Discussion}
\label{sec:disc}


Different number of quantization intervals, i.e., $Q=2, 4, 5$, have been evaluated in addition to $Q = 3$. In general, the results of analysis show a similar trend to the findings as in Table \ref{tab:patncorr}. It is observed that fewer quantization intervals may not be able to describe utterance genres of different clusters discriminatively. However, using more intervals may result in fewer salient utterance genres as restricted by the three conditions in Section \ref{ssec:expdesign}.

As a quantified representation of utterance genre, the value range has two advantages in addition to being used for automatic classification of empathy level: (1) it can provide specific range of feature values to interpret a feature pattern. For example, it is a negative sign of empathy for the therapist to speak short utterances below half of the mean duration of his or her speech utterances in a session ; (2) it can describe the utterance genre more discriminatively compared to the feature pattern. For example, in the case of two feature parameters $(p_{std}, p_{iqr})$ with $Q = 3$, it is observed that the utterance genre with the pitch range much wider than the individual baseline of the therapist could be positively correlated with empathy.

It is found that salient utterance genres often concentrate on the part of therapist utterances. For example, the average session-wise contribution ratio of salient utterance genres in the case of $Q = 3$ ranges from $6\%$ to $33\%$, with a mean of $11\%$. This suggests that the empathy level can be inferred from the part of utterances of the therapist speech in a psychotherapy session.

\section{Conclusions}
\label{sec:conc}

In this work, we propose to use utterance genres of clusters to describe the therapist's speaking style in relation to empathy in psychotherapy. The clustering is applied to utterances of therapist speech in the 118 psychotherapy sessions. The utterance genre of a cluster is represented by the quantized feature pattern of prosodic parameters. Salient utterance genres are found by analyzing the correlation between the session-wise contribution ratio of utterance genres and empathy ratings. These utterance genres are used to classify psychotherapy sessions into ``high" and ``low" empathy level, achieving an accuracy of $71\%$. The results suggest that the empathy level may be low if therapists speak long utterances slowly or speak short utterances quickly, while high if therapists talk to clients with a steady tone and volume. The observation of therapist's speaking style can be used in therapist training. For example, trainees should not only learn about what to say to clients but also the way they say it to be maximally therapeutic.

\section{Acknowledgements}
\label{sec:ack}
This research is partially supported by the Sustainable Research Fund of the Chinese University of Hong Kong (CUHK) and an ECS grant from the Hong Kong Research Grants Council (Ref.: 24604317).




\bibliographystyle{IEEEtran}

\bibliography{mybib}

\end{document}